\documentclass[aps,prd,preprint,superscriptaddress,nofootinbib]{revtex4-1}

\usepackage{latexsym}
\usepackage{amsmath}
\usepackage{amssymb}
\usepackage{amsfonts}
\usepackage{bm}

\usepackage{color}
\definecolor{purple}{rgb}{0.5,0,0.5}
\definecolor{blue}{rgb}{0.0,0,0.9}
\definecolor{prdblue}{rgb}{0.133,0.118,0.498}
\usepackage[colorlinks=true, pdfstartview=FitV, linkcolor=prdblue, citecolor= prdblue, urlcolor=prdblue]{hyperref}

\usepackage{supertabular} 
\usepackage{placeins}
\usepackage{epsfig}
\usepackage{graphicx}

\begin{document}


\title{Arising of trapped surfaces with non-trivial topology from colliding shock waves}


\author{\'Alvaro Duenas-Vidal}
\email[]{alvaro.duenas@epn.edu.ec}
\affiliation{Departamento de Física, Escuela Politécnica Nacional, Quito 170143, Ecuador}

\author{Jorge Segovia}
\email[]{jsegovia@upo.es}
\affiliation{Departamento de Sistemas F\'isicos, Qu\'imicos y Naturales, \\ Universidad Pablo de Olavide, E-41013 Sevilla, Spain}


\date{\today}

\begin{abstract}
The lightlike limit of boosted black hole solutions with one angular momentum is considered for $D \geq4$ dimensions.  The boost is performed parallel to the angular momentum and the lightlike limit is done by means of perturbative expansions.  We shown that for $D=4$ and $D> 5$ the lightlike limit cannot be extended inside the ring singularity. Then, for $D = 5$ we discuss the arising of trapped surfaces in the head-on collision. We find that, inside the validity of the perturbative analysis we do, a  trapped surface  with  topology $\mathbb R \times \mathbb S_1 \times\mathbb S_1$  seems to appear over the past light cone  of the collision below a critical value $a_c$ of the Kerr parameter $a$.
\end{abstract}


\maketitle


\section{Introduction}

The study of colliding gravitational shock waves has a long history. Previous to the attention in high energy physics, the larger class of colliding pp-wave spacetimes where thoroughly studied by purely mathematical interest. In particular, the convergence of null geodesics and production of apparent horizons after collisions where considered with certain detail~\cite{Khan:1971vh, Yurtsever:1988vc, Yurtsever:1988ks, Yurtsever:1988zz}. Once the fundamental mathematics of colliding gravitational shock waves was known, the issue was addressed within the framework of TeV-gravity as a model to estimate micro black hole cross section in high energy collisions~\cite{Yoshino:2002br, Eardley:2002re, Yoshino:2002tx}. More recently, in the framework of the AdS/CFT correspondence, colliding shock waves inside the AdS space has also been used to model thermalization after high energy collisions in the boundary field theory~\cite{Gubser:2008pc, Lin:2009pn, Lin:2010cb, Alvarez-Gaume:2008qeo, Duenas-Vidal:2010xhp, Duenas-Vidal:2012tbb}.      

The concept of gravitational waves usually refers to gravitational radiation arising from any evolving phenomena in gravitational systems~\cite{Hulse:1974eb, Weisberg:1981bh, LIGOScientific:2016aoc}. These gravitational waves are, however, weak perturbations over some stationary spacetime satisfying the Einstein field equation up to first order. Besides of gravitational radiation, there are also non-perturbative gravitational waves, as space-time wrinkles propagating over flat space satisfying \emph{exactly} the Einstein field equation for some stress-energy tensor. Gravitational shock waves belong to this second kind of gravitational waves.

Gravitational shock waves are spacetimes describing energy distributions traveling at the speed of light~\cite{Dray:1984ha, Sfetsos:1994xa}. Roughly speaking, if some energy distribution is boosted to the speed of light, the gravitational field surrounding it  squeezes by Lorentz contraction to the normal space to propagation, and that is a gravitational shock wave. In its Brinkmann form, the line element of a gravitational shock wave is
\begin{equation}\label{BrinkmannForm}
ds^2 = -dudv + d\vec{x}_\bot^2 + \delta(u)\Phi(\vec{x}_\bot) du^2 \,,
\end{equation}
where  $u = t+x, v = t-x$ are lightlike background coordinates and $\vec{x}_\bot$ spans the transverse space to the wave propagation. The function $\Phi(\vec{x}_\bot)$ gives the profile of the wave and it must satisfy the Einstein field equation for some stress-energy tensor. Note that the wave exists only at the hipersurface $u = 0$ because of the distributional term $\delta(u)$ and, after and before of the wavefront, the spacetime is flat. 

There are as many as gravitational shock waves as energy distributions we can propose to source them. For instance, a point-like sourced gravitational shock wave can be built from the lightlike limit of the boosted Schwarzschild solution. The lightlike limit must be done in a suitable way, with a subtle scaling of the mass proposed firstly by Aichelburg and Sexl~\cite{Aichelburg:1970dh}. After the lightlike limit is taken, the resulting line element can be understood as the gravitational field accompanying a massless particle~\cite{Lousto:1988ej}. The same ligthlike limit has been applied to other black-hole solutions beyond the Schwarzschild one, in order to include angular momentum and charge, but with relative success because the physical meaning of them after performing the lightlike limit is still a controversial point~\cite{Lousto:1988ua, Lousto:1989ha, Lousto:1992th, Yoshino:2004ft}. Also other shock waves, such as Giratons, have been proposed to describe polarized light beams~\cite{Frolov:2005in, Frolov:2005zq}. Finally, the lightlike limit of boosted extended black objects has been also considered in Ref.~\cite{Lousto:1990wn}.   

Since gravitational waves propagate at the speed of light, the spacetime previous to the collision of two opposite shock waves can be built by a simple \emph{linear superposition}. That is, adding a term $\delta(v)\Phi(\vec{x}_\bot) dv^2$ to the line element~\eqref{BrinkmannForm}. However, after the collision, highly non-linear gravitational interactions between the two waves take place, and the spacetime structure in the future of the collision remains unknown nowadays. Fortunately, trapped surfaces can be sought over the light cone of the collision. Then,  the appearance of  trapped surfaces  is taken as an indication that an apparent horizon forms after the collision~\cite{Giddings:2007nr, Gingrich:2007fm, Kanti:2008eq}. In an inspiring seminar at Cambridge University, Penrose showed the way a trapped surface over the past light cone of the collision can be found~\cite{penrose1974}. This trapped surface, called Penrose trapped surface from here on, is the one computed mostly in the works about colliding shock waves. Other authors, however, have considered the possibility of apparent horizons arising over the future light cone of the collision~\cite{Yoshino:2006dp, Yoshino:2007ph}. It is interesting to note that the study of colliding shock waves from boosting the Reissner-Nordström black hole shows that an apparent horizon over the future light cone can appears, while the Penrose trapped surface does not~\cite{Yoshino:2006dp, Duenas-Vidal:2012tbb}.

In this work we compute perturbatively the shock wave geometry which arises from black hole solution in $D\geq 4$ dimensions, with one angular momentum, by boosting the metric in the direction of the angular momentum until the speed of light. The ligthlike limit is taken with a mass scaling which follows the original work of Aichelburg and Sexl previously mentioned, and proposing an additional scaling for the angular momentum, such that the ring singularity of the original black hole solution is preserved through the limit. Then, the Penrose trapped surface is computed and showed that, in the scope of the perturbative method previously developed, it has ring topology for $D=5$ dimensions. Although trapped surfaces with ring topology have been found in collisions of extended objects~\cite{Yoshino:2007yk}, this is the first time that such topology appears from the head-on collision of point-like sourced shock waves.  

The manuscript is arranged as follows. In section~\ref{sec2}, a brief review of some aspect of Kerr and Myers-Perry solutions is delivered. Then, in Sec.~\ref{sec3}, we set out the lightlike limit over the boosted metric. In section~\ref{sec4} we develop a perturbative method to find the shock wave geometry inside and outside the ring singularity for $D>4$, while Sec.~\ref{sec5} is devoted to briefly discuss the case for $D=4$. Then, in Sec.~\ref{sec6}, we look for the Penrose trapped surface for $D=5$. Finally, some remarks and conclusions are given in Sec.~\ref{sec7}.


\section{Kerr-like solutions} \label{sec2}

In four dimensions, a rotating uncharged axially-symmetric black hole with a quasi-spherical event horizon can be described by the so-called Kerr solution~\cite{Kerr:1963ud}. In  Boyer-Lindsquit coordinates, the Kerr metric can be written as follows~\cite{Boyer:1966qh}:
\begin{equation}\label{k1}
ds^2 = - dt^2 + \sin^2 \theta (r^2 + a^2) d\varphi^2 + \Delta (dt - a\sin^2\theta d\varphi)^2 + \xi dr^2 + \Sigma^2 d\theta^2,
\end{equation}
with the following definitions
\begin{equation}\label{k2}
\Delta = \frac{M r}{\Sigma^2} \,, \qquad \xi = \frac{\Sigma^2}{r^2 + a^2 - rM} \,, \qquad \Sigma^2 = r^2 + a^2 \cos^2 \theta \,,
\end{equation}
where $a$ is the Kerr parameter; $M$, and $J = \frac{a M}{2G_N}$ are the mass and angular momentum of the black hole; $\Sigma$, $\Delta$ and $\xi$ are auxiliary functions.

The situation is more complicated when one goes to high dimensions. First, apart from black holes, other objects with non-spherical topologies such as black rings appear \cite{Emparan:2001wn, Emparan:2008eg}. This fact breaks the uniqueness theorems in higher dimensions. Moreover, the rotation group $SO(D-1)$ has more than one Casimir operator for $D>4$ and, therefore,  more than one angular momentum would be defined in the solutions. In particular, a $D$-dimensional rotating solutions to Einstein field equation could have up to $\llcorner (D-1)/2\lrcorner$ angular momenta. In spite of this richer taxonomy of higher dimensions, we are going to consider only the solution with spherical topology (Myers-Perry solution~\cite{perry}) and only one angular momentum. Using the same coordinate definitions than Eq.~\eqref{k1}, the mentioned solution is given by
\begin{align}\label{k3}
ds^2 &= -dt^2 +  \sin^2 \theta(r^2 + a^2) d\varphi^2 + \Delta(dt - a \sin^2\theta d\varphi)^2 + \xi dr^2 + \Sigma^2 d\theta^2 \nonumber \\
&
+ r^2 \cos^2 \theta d\Omega_{D-4}^2 \,,
\end{align}
where
\begin{equation}\label{k4}
\Delta = \frac{M}{r^{D-5}\Sigma^2} \,, \qquad \xi = \frac{r^{D-5}\Sigma^2}{r^{D-5}(r^2 + a^2) - M} \,, \qquad \Sigma^2 = r^2 + a^2 \cos^2 \theta \,.
\end{equation}
Moreover, the mass ${\cal M}$ and angular momentum ${\cal J}$ in Eq.~\eqref{k3} are related to parameters $M$ and $a$ as
\begin{equation}\label{k5}
{\cal M} = \frac{(D-2) \Omega_{D-2}}{16 \pi G_N} M \,, \qquad {\cal J} = \frac{\Omega_{D-2}}{4 \pi} J \,.
\end{equation}
The line element Eq.~\eqref{k3} is  the simplest generalization of the Kerr solution in higher dimensions. For this reason, in the following, we will refer to it as the Kerr-like metric in $D$-dimensions.  

When $a=0$, the Kerr-like metric of Eq.~\eqref{k3} reduces to the Schwarzchild metric in $D$-dimensions. Thus, in the limit $M\rightarrow 0$, we will expect the metric represents flat space-time. Setting $M=0$ in the Kerr-like metric, Eq.~\eqref{k3} reduces to
\begin{align}
ds^2 =& - dt^2 + \sin^2 \theta (r^2 + a^2) d\varphi^2 + \frac{r^2 + a^2 \cos^2 \theta}{r^2 + a^2} dr^2 \nonumber \\
&+ (r^2 + a^2 \cos^2 \theta) d\theta^2 + r^2\cos^2\theta d\Omega_{D-4}^2 \,,
\end{align}
which can be translated into the standard Cartesian form of the Minkowski metric changing to coordinates
\begin{subequations}\label{coordinates}
\begin{align}
x &= \sqrt{r^2 + a^2}\sin\theta \sin\varphi \,, \\
y &= \sqrt{r^2+ a^2}\sin\theta \cos\varphi \,, \\
z_i &= r\cos \theta \  \omega_i \,, 
\end{align}
\end{subequations}
such that $\sum_{i=1}^{D-3}\omega_i^2 = 1$. In these coordinates, the black hole is rotating in the plane $x-y$. Note that $r=0$ corresponds to the disk $x^2 + y^2 \leq a^2$ located in the plane $\sum_i z_i^2 =0$ and hence the $r$-coordinate has properties which are quite different from those of the usual radial coordinate.

The metric in Eq.~\eqref{k3} has horizons for $\xi^{-1}=0$. For $D=4$, it happens at,
\begin{equation}\label{k6}
r = \frac{M}{2}  \pm \sqrt{\frac{M^2}{4} - a^2} \,,
\end{equation}
and thus horizons appear only when $M^2 \geq 4 a^2$ is satisfied. The external horizon, $r_h = \frac{M}{2}  + \sqrt{\frac{M^2}{4} - a^2}$, will be the event horizon. For $D=5$ only one (event) horizon exists; it is located at
\begin{equation}\label{k7} 
r_h = \sqrt{M-a^2} \,,
\end{equation}
and $M>a^2$ must be satisfied. When $D>5$, it happens that
\begin{equation}\label{k8}
\lim_{r\rightarrow \infty} \xi^{-1} = 1, \qquad \lim_{r\rightarrow 0} \xi^{-1} = - \infty \,.
\end{equation}
Therefore, an event horizon always exists whatever the parameters $M$ and $a$ are for $D>5$. In all dimensions, outside the event horizon, there is a surface where $g_{tt}$ vanishes, \emph{i.e.} it changes sign inside the surface. It is located at
\begin{equation}\label{k9}
r_e(\theta) = \frac{M}{2} + \sqrt{\frac{M^2}{4}-a^2 \cos^2\theta} \,,
\end{equation}
for $D=4$, and 
\begin{equation}\label{k9}
r_e(\theta) = \sqrt{M-a^2 \cos^2\theta} \,,
\end{equation}
for $D=5$. The region $r\in (r_h, r_e)$ is called the ergosphere. 

From the expression of the metric in Eq.~\eqref{k3}, it is obvious that, besides the horizons, the metric becomes ill-defined at $\Sigma=0$. The calculation of the curvature  shows that $\Sigma=0$ is indeed a curvature singularity\footnote{In particular, for $D=4$, relative to a null tetrad based on the repeated principal null directions, the only non-zero component of the curvature tensor is $\Psi_2 = - M/[2(r+ia\cos\theta)^3]$.}. In the asymptotically Cartesian coordinates, Eqs.~\eqref{coordinates}, this corresponds to
\begin{equation}\label{singularity}
 x^2 + y^2 = a^2, \qquad \sum_i z_i^2 =0 \,.
\end{equation}
That is, the Kerr metric has a ring-like singularity of radius $a$ located in the plane $\sum_i z_i^2 =0$. 


\section{The lightlike limit} \label{sec3}

In this section we boosted the Ker-like line element Eq.~\eqref{k3} and set out the lightlike limit over it. Having as a reference the Aichelburg-Sexl limit over the Schwarzschild metric~\cite{Aichelburg:1970dh}, this implies eventually to take the limit $\gamma \rightarrow \infty$ whereas $\mu \equiv M\gamma$ remains fixed. This means that the terms of order $M^2$ and beyond in the metric will vanish after taking $\gamma \rightarrow \infty$. Therefore, it is convenient to perform a power series expansion in $M$ of the metric Eq.~\eqref{k3} and to keep just the terms of order $M^0$ and $M^1$:
\begin{equation}\label{k13}
ds^2 = ds_0^2 + M ds_1^2 + {\cal O}(M^2).
\end{equation} 
After a bit of algebra, we obtain
\begin{align}
ds_0^2 &= - dt^2 + (r^2+a^2) \sin^2\theta d\varphi^2 + \frac{r^2 + a^2 \cos^2\theta}{r^2 + a^2} dr^2\label{k14a} \nonumber \\
&
+ (r^2 + a^2 \cos^2 \theta) d\theta^2 + r^2 \cos^2\theta d\Omega_{D-4}^2 \,, \\[1ex]
ds_1^2 &= \frac{1}{r^{D-5}(r^2+ a^2 \cos^2\theta)}dt^2 - \frac{2a \sin^2 \theta}{r^{D-5} (r^2 + a^2 \cos^2 \theta)}dtd\varphi \label{k14b} \nonumber \\
&
+ \frac{a^2 \sin^4\theta}{r^{D-5} (r^2 + a^2 \cos^2 \theta}d\varphi^2 + \frac{r^2 + a^2 \cos^2 \theta}{r^{D-5} (r^2 + a^2 )^2}dr^2 \,.
\end{align} 

The boost must be done in such a way that the zero order remains unchanged. In other way, we will obtain a shock wave geometry that does not recover the flat space-time metric when $\mu \rightarrow 0$. Since $ds_0^2$ reduces to the Minkowski metric in the asymptotically Cartesian coordinates, Eqs.~\eqref{coordinates}, the authors of Ref.~\cite{Lousto:1989ha, Lousto:1992th} proposed to do the boost with respect to these coordinates. In addition, we choose to perform the boost in the direction of angular momentum. If we label this direction as $z$, the proposed coordinate transformation is
\begin{equation}\label{k18}
z = \gamma(z' + \beta t')\,, \qquad \vec{z}_\bot = \vec{z'}_\bot\,, \qquad x = x' \,,
\qquad t = \gamma(t' + \beta z') \,, \qquad y = y'.
\end{equation}
From the inverse relations to Eq.~\eqref{coordinates}
\begin{subequations}\label{k16}
\begin{align}
\cos^2 \theta &= \frac{1}{2a^2}\left[\sqrt{(\rho^2 + \vec{z}^{\ 2}- a^2)^2 + 4 a^2 \vec{z}^{\ 2}}- \rho^2 - \vec{z}^{\ 2} + a^2\right] \,, \\
r^2 &= \frac{1}{2}\left[\sqrt{(\rho^2 + \vec{z}^{\ 2}- a^2)^2 + 4 a^2 \vec{z}^{\ 2}}+ \rho^2 + \vec{z}^{\ 
2}- a^2\right] \,,
\end{align}
\end{subequations}
where $\rho^2 = x^2 + y^2$, we can compute the transformation of the components of $ds_1^2$. For $dr^2$ we have
\begin{equation} \label{k20}
dr^2 = \frac{1}{r^2 (r^2+ a^2\cos^2 \theta)^2}\left[r^2 \rho^2 d\rho^2 + (r^2 + a^2 )\vec{z}d\vec{z}\right]^2 \,.
\end{equation}
After taking the limit $\gamma\rightarrow \infty$, only the terms linear in $\gamma^2$ will survive in the transformation of $dr^2$. Thus, we can take for the transformed $dr^2$ 
\begin{equation}\label{k21}
dr^2 = \frac{(r^2 + a^2)^2}{r^2 (r^2+ a^2\cos^2 \theta)^2}\gamma^2 (z' + \beta t')^2 \gamma^2(dz' + \beta dt')^2 \,.
\end{equation}
Taking this into account and writing
\begin{equation}\label{k23}
ds_1^2 = h_{\mu\nu} dx^{\mu}dx^{\nu},
\end{equation}
the components after the boost are
\begin{subequations}\label{k24}
\begin{align}
&
h_{t't'} = \frac{\gamma^2}{r^{D-5}(r^2 + a^2 \cos^2 \theta)}\left(1 + \frac{\gamma^2 \beta^2(z'+ \beta t')^2}{r^2}\right) \,,\\
&
h_{z'z'} = \frac{\gamma^2}{r^{D-5}(r^2 + a^2 \cos^2 \theta)}\left(\beta^2 + \frac{\gamma^2 (z'+ \beta t')^2}{r^2})\right) \,,\\
&
h_{t'z'} =\frac{\gamma^2 \beta}{r^{D-5}(r^2 + a^2 \cos^2 \theta)}\left(1 + \frac{\gamma^2 (z'+ \beta t')^2}{r^2}\right) \,,\\ 
&
h_{\varphi' \varphi'} = \frac{a^2 \sin^4 \theta}{r^{D-5} (r^2 + a^2 \cos^2 \theta)} \,,\\
&
h_{t' \varphi'} = \frac{\gamma a \sin^2 \theta}{r^{D-5} (r^2 + a^2 \cos^2 \theta)} \,,
\end{align}
\end{subequations}
where
\begin{equation}\label{k22}
r^2 + a^2 \cos^2\theta = \sqrt{({\rho'}^2 + \gamma^2(z' + \beta t')^2+ \vec{z'}_\bot^{\ 2}- a^2)^2 +4 a^2 \gamma^2 (z' + \beta t')^2+ 4 a^2 \vec{z'}_\bot^{\ 2}} \,.
\end{equation}

A priori it seems that the components of Eqs.~\eqref{k24} grow monotonically without limit when $\gamma$ approaches $\infty$. In order to \emph{regularize} this behaviour, $\mu =\gamma M$ is fixed while taking $\gamma \rightarrow \infty$; as in Ref.~\cite{Lousto:1989ha, Lousto:1992th}, we also keep $a$ fixed throughout the process. In this way, we obtain
\begin{equation}
h_{\varphi'\varphi'}=0, \qquad h_{t'\varphi'} =0.
\end{equation}
The remaining components, in lightlike background coordinates $u' = z'+t'$ and $v' = z'-t'$, are
\begin{subequations}\label{k26}
\begin{align}
&
h_{u'u'} = \frac{\gamma^2}{4 r^{D-5}(r^2 + a^2 \cos^2 \theta)} (1+\beta )^2 \left(1 + \frac{\gamma^2(z'+ \beta t')^2}{r^2}\right) \,, \\
&
h_{v'v'} = \frac{\gamma^2}{4 r^{D-5}(r^2 + a^2 \cos^2 \theta)} (1-\beta )^2 \left(1 + \frac{\gamma^2(z'+ \beta t')^2}{r^2}\right) \,, \\
&
h_{u'v'} =  \frac{\gamma^2}{4 r^{D-5}(r^2 + a^2 \cos^2 \theta)} (\beta^2 -1) \left(1 - \frac{\gamma^2(z'+ \beta t')^2}{r^2}\right) \,.
\end{align}
\end{subequations}
Thus, only the component $h_{u'u'}$ survives to the imposed limit. 

Summarizing, from Eq.~\eqref{k14b} and the $(u'u')$-component of Eq.~\eqref{k26}, the lightlike limit over the line element Eq.~\eqref{k13} is 
\begin{equation}\label{k27}
ds^2 = -dudv + dx^2 + dy^2 + d\vec{z}_\bot^{\ 2} + F(t,\rho^2,z, R^2) du^2 \,,
\end{equation}
which is the metric of a shock wave. The primed coordinates have been changed by unprimed ones for clarity, $R^2$ stands for $\vec{z}_\bot^{\ 2}$, and $F(t, \rho^2,z,  R^2)$ is  given by the limit
\begin{equation}\label{k28}
F(t,\rho^2,z, R^2)= \mu \lim_{\gamma\rightarrow \infty} \gamma f\left(\gamma^2 (z+ \beta t)^2, \rho^2, R^2\right), 
\end{equation}
being $f(w^2, \rho^2, R^2)$ the function
\begin{align}\label{k28b}
f(w^2, \rho^2, R^2) &= \frac{2^{\frac{D-5}{2}}}{\left[\sqrt{(\rho^2 + w^2 + R^2- a^2)^2 + 4a^2 (w^2 + R^2)}+ \rho^2 + w^2 + R^2- a^2\right]^\frac{D-5}{2}} \nonumber \\
& \times \frac{1}{ \sqrt{(\rho^2 + w^2 + R^2- a^2)^2 + 4a^2 (w^2 + R^2)}} \nonumber \\
& \times\left[1 + \frac{2w^2}{\sqrt{(\rho^2 + w^2 + R^2- a^2)^2 + 4a^2 (w^2 + R^2)}+ \rho^2 + w^2 + R^2- a^2}\right].
\end{align}

It is worth noting herein that we have obtained a shock wave from a rotating solution; however, the following limit has been imposed
\begin{equation}\label{k28c}
\lim_{\gamma \rightarrow \infty} J \propto \lim_{\gamma \rightarrow \infty} M a =0 \,,
\end{equation}
since we have fixed $\mu =M\gamma$ and $a$. This means that the shock wave has no angular momentum. From a physical point of view, it is not surprising given that any observer can not measure rotation in an ultrarelativistic object. However, this lack of angular momentum has been used as an argument to reject the shock wave of Eq.~\eqref{k27} as a valid model for ultrarelativistic energy lumps with spin. Gyratons has been proposed instead but, strictly speaking, a rest solution for gyratons is not known and thus they cannot work as a model to describe high energy collision between heavy ions. In addition, note that in both cases the angular momentum is of orbital type, and not spin. Finally, in Ref.~\cite{Burinskii:1999ew} a shock wave with angular momentum has been obtained from a Kerr solution, but a re-scaling of $a$ is mandatory. Since $a$ measures distances in the plane $xy$, the re-scaling is equivalent to change distances in the plane perpendicular to the boost direction, which does not seem a good idea from a physical point of view.  


\section{Perturbative computation for $D>4$}\label{sec4}

We turn now our attention to the computation of the limit in Eq.~\eqref{k28} for $D>4$. The way to proceed is basically finding a primitive of $\gamma f\left(\gamma^2 (z+ \beta t)^2, \rho^2, R^2\right)$ with respect to $z$, compute the limit, and derive it. When $f(w^2, \rho^2, R^2)$ is integrable with respect to $w$, this method is resumed in the equation
\begin{equation}\label{lemma}
\lim_{\gamma\rightarrow \infty} \gamma f\left(\gamma^2 (z+ \beta t)^2, \rho^2, R^2\right) = \delta(u) \int_{-\infty}^{\infty} dw f(w^2, \rho^2, R^2). 
\end{equation}
In any case, given that the Kerr-like metric has a ring singularity located at $x^2 + y^2 = a^2$, $\sum_i z_i^2 =0$, it is necessary to compute the limit separately for $\rho^2<a^2$ (interior solution) and $\rho^2>a^2$ (exterior solution).

\subsection{Interior solution}

The integral in the right part of Eq.~\eqref{lemma} is hard to compute because the very complicated form of the function $f(w^2, \rho^2, R^2)$ in Eq.~\eqref{k28b}. To deal with this problem, we define dimensionless variables
\begin{equation}\label{k29}
 \bar w = \frac{w}{a}, \qquad \bar \rho = \frac{\rho}{a}, \qquad \bar R = \frac{R}{a},
\end{equation}
and propose to do an expansion of $f(w^2, \rho^2, R^2)$ in powers of $\bar \rho$ to integrate order by order.\footnote{Inside the ring singularity we have $\bar \rho  < 1$ and thus the expansion, as well as the integration, makes sense.}

After some algebraic manipulations, we get the following expression
\begin{equation}\label{k30}
a^{D-4} f(\bar w^2, \bar \rho^2, \bar R^2) = \sum_{n=0}^\infty I_{2n}^{(D)}(\bar w^2 , \bar R^2) \bar \rho^{2n} + 2 \bar w^2 \sum_{n=0}^\infty I_{2n}^{(D+2)}(\bar w^2 , \bar R^2) \bar \rho^{2n} \,,
\end{equation}
where
\begin{equation}\label{k31}
I_{2n}^{(D)}(\bar w^2, \bar R^2) = \frac{(-1)^n}{2^n n! (\bar w^2 + \bar R^2)^{\frac{D-5}{2}} (\bar w^2 +\bar R^2 +1)^{2n+1}} \sum_{k=0}^{n}\binom{n}{k} P_n^{(k)}(D) (\bar w^2 + \bar R^2)^k \,,
\end{equation}
being $P_n^{(k)}(D)$ polynomials in $D$ of degree $n$, defined as\footnote{The $(a)^{n,k}$ is the generalized descendent Pochhammer symbol, defined as $(a)^{n,k} = a (a-k)(a-2k)(a-3k)\ldots (a-(n-1)k)$ and $(a)^{0,k} =1$.}
\begin{equation}\label{k32}
P_n^{(k)}(D) = (D-5+2n)^{k,2} (D-7-2k)^{n-k,2} \,.
\end{equation}

The integration of the first term in Eq.~\eqref{k30} gives
\begin{align}\label{k33}
\int_{-\infty}^\infty dw I_{2n}^{(D)}(\bar w^2 , \bar R^2) &= \frac{(-1)^n}{2^n n!}\frac{\bar w}{\bar R^{D-5} (1 + \bar R^2)^{2n+1}} \sum_{k=0}^{n}\binom{n}{k} P_n^{(k)}(D) \nonumber \\
&
\times \bar R^{2k} \left.\mathfrak F\left(\frac{1}{2}; \frac{D-5-2k}{2}, 2n+1; \frac{3}{2}; \frac{-\bar w^2}{\bar R^2}, \frac{-\bar w^2}{1 +\bar R^2}\right)\right|_{-\infty}^{\infty} \,.
\end{align}
Using the properties
\begin{align}\label{k34}
&  \mathfrak F(a;b_1, b_2; c; z_1, z_2) = (1-z_2)^{-a} \mathfrak F\left(a;b_1, c-b_1-b_2; c; \frac{z_2-z_1}{z_2-1}, \frac{z_2}{z_2-1}\right) \,, \\
& \mathfrak F(a;b_1, b_2; c; z_1, 1) = {\ }_2F_1(a,b_2;c;1) {\ }_2F_1(a,b_1,c-b_2;z_1) \,,
\end{align}
together with the asymptotic value
\begin{equation}\label{k35}
{\ }_2F_1(a,b;c;1) = \frac{\Gamma(c)\Gamma(c-a-b)}{\Gamma(c-a)\Gamma(c-b)} \,,
\end{equation}
we can evaluate Eq.~\eqref{k33}, obtaining
\begin{align}\label{k36}
\int_{-\infty}^\infty dw I_{2n}^{(D)}(\bar w^2 , \bar R^2) &= \frac{(-1)^n\sqrt{\pi}}{2^n n! \bar R^{D-5} (1 + \bar R^2)^{\frac{4n+1}{2}}} \sum_{k=0}^{n}\binom{n}{k} P_n^{(k)}(D) \bar R^{2k} \nonumber \\
&
\times \frac{\Gamma\left(\frac{D-4+4n-2k}{2}\right)}{\Gamma\left(\frac{D-3+4n-2k}{2}\right)} {\ }_2F_1 \left(\frac{1}{2}, \frac{D-5-2k}{2}; \frac{D-3+4n-2k}{2}; -\frac{1}{\bar R^2}\right) \,.
\end{align}

Proceeding in a similar way, the integration of the second term in Eq.~\eqref{k20} gives
\begin{align}\label{k37}
\int_{-\infty}^\infty dw 2 \bar w^2 & I_{2n}^{(D+2)}(\bar w^2 , \bar R^2) = \frac{2\sqrt{\pi}(-1)^n }{2^{n} n! \bar R^{D-3} (1 + \bar R^2)^{\frac{4n-1}{2}}} \sum_{k=0}^{n}\binom{n}{k} P_n^{(k)}(D+2) \nonumber \\ 
&
\times\bar  R^{2k}\left[ \frac{\Gamma\left(\frac{D-4+4n-2k}{2}\right)}{\Gamma\left(\frac{D-3+4n-2k}{2}\right)} {\ }_2F_1 \left(\frac{1}{2}, \frac{D-3-2k}{2}; \frac{D-3+4n-2k}{2}; -\frac{1}{\bar R^2}\right) \right. \nonumber \\
&
\left.- \frac{\Gamma\left(\frac{D-2+4n-2k}{2}\right)}{\Gamma\left(\frac{D-1+4n-2k}{2}\right)} {\ }_2F_1 \left(\frac{1}{2}, \frac{D-3-2k}{2}; \frac{D-1+4n-2k}{2}; -\frac{1}{\bar R^2}\right) \right].
\end{align}

Therefore, from Eqs.~\eqref{lemma}, \eqref{k30}, \eqref{k36} and \eqref{k37}, we have finally for the interior solution
\begin{equation}\label{k38}
 F(t,\rho^2,z, R^2)= \delta(u) \Phi_-(\bar\rho^2, R^2),
\end{equation}
where
\begin{align}\label{interior}
\Phi_-(\bar \rho^2, R^2) &= \frac{\mu \sqrt{\pi}}{a^{D-4}}\sum_{n=0}^{\infty}\frac{(-\bar \rho^2)^n}{2^n n!\bar  R^{D-5} (1 + \bar R^2)^{\frac{4n+1}{2}}} \sum_{k=0}^{n}\binom{n}{k} P_n^{(k)}(D) \bar R^{2k} \nonumber \\
&
\times \frac{\Gamma\left(\frac{D-4+4n-2k}{2}\right)}{\Gamma\left(\frac{D-3+4n-2k}{2}\right)} {\ }_2F_1 \left(\frac{1}{2}, \frac{D-5-2k}{2}; \frac{D-3+4n-2k}{2}; -\frac{1}{\bar R^2}\right) \nonumber \\
&
+ \frac{2\mu \sqrt{\pi}}{a^{D-4}}\sum_{n=0}^{\infty} \frac{(-\bar \rho^2)^n }{2^{n} n! \bar R^{D-3} (1 + \bar R^2)^{\frac{4n-1}{2}}} \sum_{k=0}^{n}\binom{n}{k} P_n^{(k)}(D+2) \nonumber \\ 
&
\times\bar  R^{2k}\left[ \frac{\Gamma\left(\frac{D-4+4n-2k}{2}\right)}{\Gamma\left(\frac{D-3+4n-2k}{2}\right)} {\ }_2F_1 \left(\frac{1}{2}, \frac{D-3-2k}{2}; \frac{D-3+4n-2k}{2}; -\frac{1}{\bar R^2}\right) \right. \nonumber \\
&
\left.- \frac{\Gamma\left(\frac{D-2+4n-2k}{2}\right)}{\Gamma\left(\frac{D-1+4n-2k}{2}\right)} {\ }_2F_1 \left(\frac{1}{2}, \frac{D-3-2k}{2}; \frac{D-1+4n-2k}{2}; -\frac{1}{\bar R^2}\right) \right] \,.
\end{align}

Note that this general result holds only for $\bar R^2\neq 0$. When $\bar R^{2} = 0$, each term in Eq.~\eqref{k30} takes the form
\begin{align}\label{k39}
I_{2n}^{(D)}(\bar w^2 , \bar R^2) &+ 2 \bar w^2 I_{2n}^{(D+2)}(\bar w^2 , \bar R^2) \nonumber \\
& 
= \frac{(-1)^n}{2^n n! |w|^{D-5} (1+w^2)^{2n+1}}\sum_{k=0}^n\binom{n}{k} \left[P_n^{(k)}(D) + 2 P_n^{(k)}(D+2)\right] \,.
\end{align}
Therefore, when $D>5$, negative powers in $|w|$ appear in the expansion of $f(\bar w^2, \bar \rho^2, \bar R^2=0)$, and thus it is not integrable in the plane $\bar R^2 =0$. It has been argued in Ref.~\cite{Yoshino:2004ft} that this implies the disk $x^2 + y^2 \leq a^2$, $R=0$ is a curvature singularity. However, strictly speaking, the fact that Eq.~\eqref{k30} is not integrable in $R^2=0$ only tells us that we can not use the relation \eqref{lemma} to compute the limit because such relation only applies to integrable functions. It could happen that the lightlike limit Eq.~\eqref{k28} was well defined in $R=0$; the correct way to check it is to see if \eqref{interior} can be extended analytically to $R=0$. A fast inspection in this sense shows that only for $D=5$ the solution can be extended into the ring.

\subsection{Exterior solution}

We are going to solve the integral of Eq.~\eqref{lemma} beyond the ring singularity. In order to do so, we redefine the coordinate $\rho$ as
\begin{equation}
\rho^2-a^2 \rightarrow \rho^2 \,,
\end{equation}
and perform an expansion in powers of $a^2$ of the function $f(w^2, \rho^2, R^2)$,
\begin{equation}\label{k40}
f(w^2, \rho^2, R^2) = \sum_{n=0}^\infty J_{2n}^{(D)}(w^2, \rho^2, R^2) a^{2n} \,,
\end{equation}
to integrate order by order. In this way we are studying how different is the shock wave obtained form the Kerr-like line element from the one generated boosting the Schwarzschild solution,  out of the region bounded by the cylinder $x^2 + y^2 = a^2$.  

After some algebraic manipulations, we get
\begin{align}\label{k41}
J_{2n}^{(D)}(w^2, \rho^2, R^2) &= \frac{(-1)^n}{2^n n!} R_{n-1}(D) \nonumber \\
&\times\frac{(w^2 + R^2)^n\left[2(D+ 3n-3) w^2 + (D-3 + 2n)(R^2 + \rho^2)\right]}{(w^2 + R^2 + \rho^2)^{\frac{D+4n-1}{2}}} \,,
\end{align}
where $R_n(D)$ are polynomials in $D$ of degree $n$, defined as\footnote{The $(b)_{n,k}$ is the generalized ascendant Pochhammer symbol, defined as $(b)_{n,k} = b (b+k)(b+2k)(b+3k)\ldots (b+(n-1)k)$ and $(b)_{0,k} =1$.}.
\begin{equation}\label{k42}
 R_n(D) = \left(D-3+2(n-1)\right)_{n,2}, \qquad R_{-1}(D) = \frac{1}{D-3}.
\end{equation}
The integration of the general term $J_{2n}^{(D)}$ gives
\begin{align}\label{k43}
\int_{-\infty}^\infty dw & J_{2n}^{(D)}(w^2, \rho^2, R^2) = \frac{(-1)^n}{2^n n!}R_{n-1}(D) \frac{w\ R^{2n}}{(R^2 + \rho^2)^{\frac{D+4n-1}{2}}} \nonumber \\
&\times\left[ 3(D+2n-3) (R^2 + \rho^2) \mathfrak{F} \left(\frac{1}{2}, -n, \frac{D+4n-1}{2}, \frac{3}{2}, -\frac{w^2}{R^2}, -\frac{w^2}{R^2 + \rho^2}\right) \right. \nonumber \\
&\left. \left. + 2(D+3n-3) w^2 \mathfrak{F}\left( \frac{3}{2}, -n, \frac{D+4n-1}{2}, \frac{5}{2}, -\frac{w^2}{R^2}, -\frac{w^2}{R^2 + \rho^2}\right) \right]\right|_{w \rightarrow-\infty}^{w \rightarrow \infty}
\end{align}
To evaluate it we use one more time the relations \eqref{k34} and \eqref{k35}. Finally we obtain,
\begin{align}\label{k44}
&
\int_{-\infty}^\infty dw J_{2n}^{(D)}(w^2, \rho^2, R^2) = 6 \frac{(-1)^n}{2^n n!}R_{n-1}(D) \frac{ R^{2n}}{(R^2 + \rho^2)^{\frac{D+4n-4}{2}}} \nonumber \\
&
\times \left[ {\ }_2F_1\left( \frac{1}{2}, -n, \frac{D+2n-1}{2}, -\frac{\rho^2}{R^2}\right) + \frac{2(D+3n-3)}{D+2n-3} {\ }_2F_1\left(\frac{3}{2}, -n, \frac{D+2n-1}{2}, -\frac{\rho^2}{R^2}\right)\right] \,.
\end{align}
Therefore, writing the exterior solution in Eq.~\eqref{k27} as
\begin{equation}\label{k45}
F(t,\rho^2, z, R^2) = \delta(u) \Phi_+(\rho^2, R^2) \,,
\end{equation}
from Eqs.~\eqref{lemma}, \eqref{k40} and \eqref{k44}, we have
\begin{align} \label{k46}
\Phi_+(\rho^2, R^2) &=  6\mu\sum_{n=0}^\infty \frac{(-a^2)^n}{2^n n!}R_{n-1}(D) \frac{ R^{2n}}{(R^2 + \rho^2)^{\frac{D+4n-4}{2}}} \nonumber \\
&
\times \left[ {\ }_2F_1\left( \frac{1}{2}, -n, \frac{D+2n-1}{2}, -\frac{\rho^2}{R^2}\right)\right. \nonumber \\
&
\left. + \frac{2(D+3n-3)}{D+2n-3} {\ }_2F_1\left(\frac{3}{2}, -n, \frac{D+2n-1}{2}, -\frac{\rho^2}{R^2}\right)\right] \,.
\end{align}

\begin{figure*}[!t]
\centering
\includegraphics[width=0.95\textwidth, height=0.33\textheight]{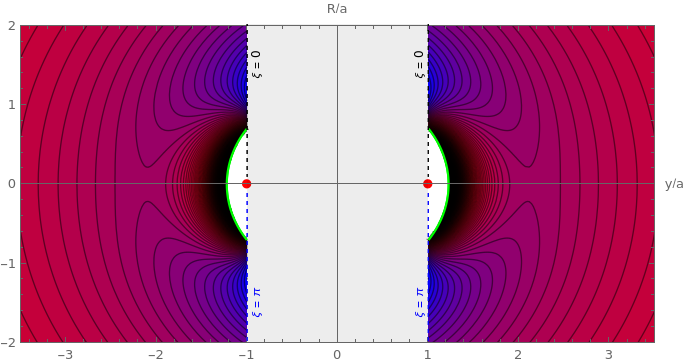}
\caption{\label{fig1} Contour graph of the exterior solution Eq.~\eqref{k46} over the slice $x=0$ up to first order. Red points mark the location of the ring singularity $x^2 + y^2 = a^2, R^2 = 0$. The region in white, delimited by the green arcs,  corresponds to the region $x^2 + y^2 + R^2 \leq 3/2$, where the convergence of Eq.~\eqref{k46} can not be assured. Finally, the shadowed region in gray color corresponds to the domain of the interior solution, Eq.~\eqref{interior}, which is not plotted for clarity.}
\end{figure*}

It is convenient to introduce new coordinates $\{\eta, \xi,\chi, \vartheta_i\}$ given by
\begin{subequations}\label{k47}
\begin{align}
x &= (\eta^2 \sin^2\xi + a^2)^{\frac{1}{2}} \sin{\chi} \,, \\
y &= (\eta^2 \sin^2\xi + a^2)^{\frac{1}{2}} \cos{\chi} \,, \\
{z_\bot}_i &= \eta \cos\xi \vartheta_i \,,
\end{align}
\end{subequations}
such that $\sum_{i=1}^{D-4} \vartheta_i^2 = 1$, $\eta \in \mathbb R^+$, $\xi\in[0, \pi]$ and $\xi \in [0, 2\pi)$.  With these coordinates, $\Phi_+$ reads
\begin{align}\label{k48}
\Phi_+(\eta, \xi) &= \frac{6\mu}{\eta^{D-4}}\sum_{n=0}^\infty \left(-\frac{a^2}{2 \eta^2}\right)^n\frac{R_{n-1}(D) \cos^{2n}\xi}{ n!} \nonumber \\
&
\times\left[ {\ }_2F_1\left( \frac{1}{2}, -n, \frac{D+2n-1}{2}, -\tan^2\xi \right)\right. \nonumber \\
&
\left. \ \  + \frac{2(D+3n-3)}{D+2n-3} {\ }_2F_1\left(\frac{3}{2}, -n, \frac{D+2n-1}{2}, -\tan^2\xi\right)\right] \,.
\end{align}
Note that this is a multipole expansion of $\Phi_+$ in powers of $a^2/2\eta^2$ outside the region delimited by  the surface $x^2 + y^2 = a^2$. Therefore we can assure the convergence of Eq.~\eqref{k48} whenever $a^2 /2\eta^2 < 1$. Since, from Eqs.~\eqref{k47}
\begin{equation}\label{PerturbativeCond}
 \left(\frac{\eta}{a}\right)^2- \left(\frac{R}{a}\right)^2 = \frac{x^2 + y^2}{a^2} -1\, ,
\end{equation}
we cannot assure the convergence of the perturbative expansion Eq.~\eqref{k48}  inside the region 
\begin{equation}\label{validityExtSol}
x^2 + y^2 + R^2 \leq  \frac{3}{2} a^2\,, \quad x^2 + y^2 \geq a^2.
\end{equation}
In Fig. \ref{fig1} a graph of the perturbative expansion  Eq.~\eqref{k48} until the mode $n = 1$ is shown together with the region where the convergence of Eq.~\eqref{k48} might be compromised. 

\section{Brief review of the shock wave in $D=4$}\label{sec5}

The construction of the shock wave fro $D=4$ is delicate because the lack of degrees of freedom, since the variable $R$ does not exist in four dimensions. In this way, the integral of the terms in Eq.~\eqref{k30} is reduced to
\begin{align}\label{k49}
\int_{-\infty}^{\infty}d\bar w & \left[ I_{2n}^{(4)}(\bar w^2 , \bar R^2) + 2 \bar w^2 I_{2n}^{(D+2)}(\bar w^2 , \bar R^2)\right] = \frac{(-1)^n}{2^n n!} \nonumber \\
&
\times \sum_{k=0}^n\binom{n}{k} \left[P_n^{(k)}(4) + 2 P_n^{(k)}(6)\right]\int_{-\infty}^{\infty}\frac{d\bar w |\bar w|}{ (1+\bar w^2)^{2n+1}} \,.
\end{align}
The mode $n=0$ gives a non finite result:
\begin{equation}\label{k50}
\int_{-\infty}^{\infty}\frac{d\bar w |\bar w|}{ (1+\bar w^2)} = 2  \int_{0}^{\infty}\frac{d\bar w \bar w}{ (1+\bar w^2)} = \left.\log (1+\bar w^2)\right|_{0}^\infty \rightarrow \infty \,.
\end{equation}
Therefore, it is not clear if an interior solution is possible in four dimensions. In fact, it may not exist since such solution should satisfy some Einstein equations but the ring singularity forbids any boundary condition. We can only state that, from this point of view, the problem is not well posed.  

On the other hand, the exterior solution can be computed, as it has been shown in Ref.~\cite{Lousto:1989ha}. Note that in this case the solution \eqref{k46} is valid except to the order $n=0$, since the integral \eqref{k43} is not finite for $n=0$ and $D=4$. In Ref.~\cite{Lousto:1988ua} a procedure to regularize the limit has been defined. Given that for $n=0$ we are faced with the  shock wave arising from boosting the Schwarzschild black hole, we can use the results of Ref.~\cite{Duenas-Vidal:2010xhp} without computing again. Then we have
\begin{equation}\label{k51}
\lim_{\gamma\rightarrow \infty} \gamma J_0^{(4)}\left(\gamma^2 (z+\beta t)^2, \rho^2, R^2\right) = -2 \delta(u)\log{(\rho^2+ R^2)}+ \frac{2}{|u|} \,.
\end{equation}

 
\section{Head-on collision} \label{sec6}

Let us assume that two shock waves collide at $t=0$ with zero impact parameter, and such that their profiles are given by Eqs.~\eqref{interior} and~\eqref{k48}. Then, outside the chronological future of the collision surface, the metric is given by
\begin{equation}\label{k52}
ds^2 = -dudv + dx^2 + dy^2 + d\vec{z}^{\ 2}_\bot + \delta(u) \Phi(\bar \rho^2,\bar R^2)du^2 + \delta(v)\Phi(\bar \rho,\bar R^2)dv^2 \,,
\end{equation}
where $\Phi = \Phi_+(\eta,\xi)$ ($\Phi = \Phi_-(\bar\rho^2, R^2)$) whenever we are outside (inside) the region $x^2 + y^2 = a^2$.

From the line element Eq.~\eqref{k52}, now we  look for the Penrose trapped surface. It is a marginally outer trapped surface  lying in the past light cone of the collision, i.e. a spacelike $(D-2)$ surface inside the region $\{u\leq 0, v = 0\}\bigcup\{u = 0, v\leq 0\}$ whose outer null normals have zero convergence. Note that because $a\ne0$, the rotation group $SO(D-2)$ that acts over the wavefront is broken into $SO(2) \times SO(D-4)$; this signals the possibility of torus topology for the Penrose trapped surface. To get the equations satisfied by the Penrose trapped surface it is necessary to choose suitable coordinates $ \{U,V,X,Y, \vec{Z}_\bot\}$ such that the null geodesics normal to the wavefronts are continuous. Then, parametrizing the Penrose surface $\mathcal S = \mathcal S_u \bigcup \mathcal S_v$ by a function $\Psi(X,Y,\vec{Z}_\bot) \geq 0$  as 
\begin{subequations}\label{PenroseSurface}
\begin{align}
\mathcal S_u&= \left\{(U,V, X,Y,\vec{Z}_\bot)\ :\ U = 0,\ V+ \Psi = 0\right\}\, , \\
\mathcal S_v &= \left\{(U,V, X,Y,\vec{Z}_\bot)\ :\ U + \Psi= 0,\ V=0 \right\}\, ,
\end{align}
\end{subequations}
to find the Penrose trapped surface is equivalent to solve the boundary problem given by (see Refs.~\cite{Eardley:2002re, Yoshino:2002tx, Lin:2010cb, Alvarez-Gaume:2008qeo} for details):
\begin{subequations}\label{k54}
\begin{align}
\bigtriangleup_\bot(\Phi-\Psi) &= 0 \,, \\
\left.\Psi\right|_{\mathcal C} &= 0 \,, \\
\left. g_\bot^{ab}\partial_a\Psi\partial_b\Psi\right|_{\mathcal C} &= 4 \,,
\end{align}
\end{subequations}
where $\mathcal C$ is the intersection of the Penrose trapped surface with the collision surface $u = v = 0$. Given that we have a geometry split by the cylinder $x^2 + y^2 = a^2$, we are forced to compute separately the exterior and interior pieces of the trapped surface $\mathcal S$.

\subsection{Exterior trapped surface for $D>4$}

Let $\Psi_+(\eta,\xi)$ the function parameterizing the piece of the Penrose trapped surface which is exterior to the cylinder $x^2 + y^2 = a^2$. The solution to the first equation in~\eqref{k54} with the first boundary condition is given by
\begin{equation}\label{k55}
\Psi_+(\eta,\xi) = \Phi_+(\eta,\xi) - \Phi_+(\eta(\xi), \xi) \,,
\end{equation}
being $\eta=\eta(\xi)$ a parameterization of $\mathcal C$. Thus the second boundary condition reads
\begin{equation}\label{k56}
\frac{a^2 + \eta^2}{\eta^2}\left[\partial\Psi_+(\eta(\xi),\xi)\right]^2 = 4 \,,
\end{equation}
where we have used the fact that $\partial_\xi\Psi_+(\eta(\xi),\xi)=0$ because $\Psi_+(\eta(\xi), \xi) =0$. On the other hand, since $\Psi_+$ is defined such that $\Psi_+ > 0$ inside $\mathcal C$ and $\Psi_+=0$ in $\mathcal C$, one must have $\partial_\eta\Psi_+|_{\mathcal C} <0$. Thus, the later equation reduces to
\begin{equation}\label{k57}
\partial\Psi_+(\eta(\xi),\xi) = -2 \frac{\eta}{\sqrt{\eta^2+a^2}} \,. 
\end{equation}
Substituting here the solution~\eqref{k55}, it gives the algebraic equation
\begin{align}\label{k58}
&
\mu \frac{(D-2) (D-4)\sqrt{\pi}\Gamma\left(\frac{D-4}{2}\right)}{(D-3)\Gamma\left(\frac{D-3}{2}\right)} + 6\mu \sum_{n=1}^\infty \left(-\frac{a^2}{\eta^2}\right)^n \nonumber \\
&
\times\frac{P_n(D) (D+2n-4)\cos^{2n}\xi}{(2n)!!}\left[ {\ }_2F_1\left( \frac{1}{2}, -n, \frac{D+2n-1}{2}, -\tan^2\xi \right)\right. \nonumber \\
&
\left. + \frac{2(D+3n-3)}{D+2n-3} {\ }_2F_1\left(\frac{3}{2}, -n, \frac{D+2n-1}{2}, -\tan^2\xi\right)\right] \nonumber \\
&= 2\eta^{D-3} \frac{1}{\sqrt{1 + \frac{a^2}{\eta^2}}} \,.
\end{align}

To solve the last equation, we assume first a series expansion of $\eta(\xi)$ in powers of $a^2$, 
\begin{equation}\label{k59}
\eta(\xi) = \eta_0(\xi) + \eta_2(\xi) a^2 + \frac{1}{2}\eta_4(\xi) a^4 + \ldots
\end{equation}
second substituting it in Eq.~\eqref{k44}, and solve order by order in $a^2$. At zero order, we have the solution
\begin{equation}\label{k60}
\eta_0^{D-3} = \mu \frac{(D-2) (D-4)\sqrt{\pi}\Gamma\left(\frac{D-4}{2}\right)}{2(D-3)\Gamma\left(\frac{D-3}{2}\right)}.
\end{equation}
Thus $\eta_0$ is a constant. At second order, we obtain
\begin{align}\label{k61}
\eta_2(\xi) =& \frac{1}{2 (D-3) \eta_0}- 6\mu \frac{(D-2)}{2(D-3)\eta_0^{D-2}}\cos^2\xi \nonumber \\
&
\times \left[{\ }_2F_1\left(\frac{1}{2}, -1, \frac{D+1}{2}, - \tan^2 \xi\right) \right. \nonumber \\
&
\hspace{1cm}\left.+ \frac{2D}{D-1}{\ }_2F_1\left(\frac{3}{2}, -1, \frac{D+1}{2}, - \tan^2 \xi\right)\right] \,.
\end{align}
Note that the hypergeometrical function defined as
\begin{subequations}\label{k62}
\begin{align}
 &{\ }_2F_1(a,b,c,z) \equiv\sum_{n=0}^\infty\frac{(a)_n (b)_n}{(c)_n}\frac{z^n}{n!}, \\
&(a)_n = a(a+1)(a+2)\ldots (a+n-1),
\end{align}
\end{subequations}
is equal to a polinomial of finite order when either $a$ or $b$ are negative integers. In particular,
\begin{equation}\label{k63}
\begin{split}
&{\ }_2F_1\left(\frac{1}{2}, -1, \frac{D+1}{2}, - \tan^2 \xi\right) = 1 + \frac{\tan^2 \xi}{D+1},\\
&{\ }_2F_1\left(\frac{3}{2}, -1, \frac{D+1}{2}, - \tan^2 \xi\right) = 1 + \frac{3\tan^2 \xi}{D+1}.
\end{split}
\end{equation}
Then, Eq.~\eqref{k61} is simplified to
\begin{equation}\label{k64b}
\eta_2(\xi) = \frac{1}{2 (D-3) \eta_0}- 6 \mu \frac{(D-2)}{2(D-3)\eta_0^{D-2}}\cos^2\xi \left(3D-1 + \frac{7D-1}{D+1}\tan^2\xi\right) \,.
\end{equation}

\begin{figure*}[!t]
\centering
\begin{tabular}{lll}
\includegraphics[width=0.45\textwidth, height=0.30\textheight]{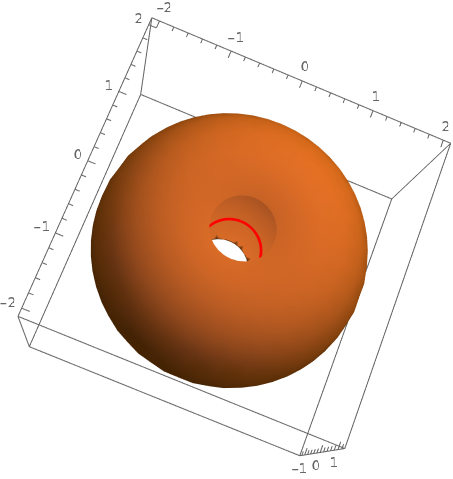}
&
\hspace*{0.60cm}
&
\includegraphics[width=0.45\textwidth, height=0.30\textheight]{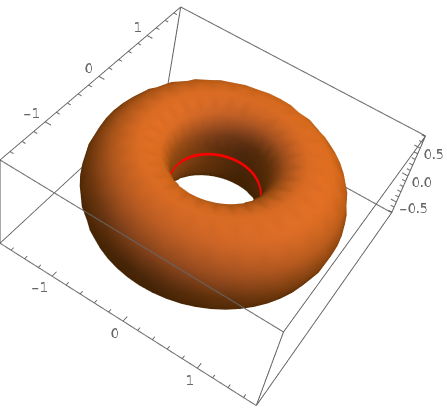}
\end{tabular}
\caption{\label{fig2} The shape of the piece of $\mathcal C$ outside of the cylinder $x^2 + y^2=a^2$ for $D=5$ with energy $\mu = 1$. The ring singularity is plotted in red. \emph{Left panel}.- For a value $a<a_0$, the piece of $\mathcal C$ outside the cylinder $x^2 + y^2 = a^2$ is not closed and must continue inside $x^2 + y^2 = a^2$. \emph{Right panel}.- For $a = a_0$ the surface $\mathcal C$  closes entirely outside $x^2 + y^2 = a^2$.}
\end{figure*}

By definition $\eta \geq 0$ and it becomes zero just over the ring singularity $x^2 + y^2 = a^2$, $R=0$. However, there are values of $\mu$ and $a$ so that $\eta(\xi) = \eta_0 + \eta_2(\xi) a^2 + O(a^4)$ may take negative values. This indicates that, for each energy $\mu$, there is a  value $a_0$ such that we can be sure there exists no solution to the trapped surface equations for $a^2>a^2_0$. To determine this value, we impose $\eta(\xi = 0) = 0$. Solving this algebraic equation for $D=5$,\footnote{Although the outer piece of the trapped surface can be computed for $D>5$ whit the perturbative method we develop in Sec.~\ref{sec4}, the inner one only has sense for $D=5$ because the interior geometry is singular in other dimensions.} up to order $a^2$, the value of $a_0$ is
\begin{equation}\label{k64}
 a_0^2 =\frac{6 \pi^2 \mu}{84  -\pi}\simeq 0.73\mu.
\end{equation}
When $a^2=a^2_0$, the exterior piece of $\mathcal C$ is closed, and the ring singularity is over it. For $a^2<a^2_0$ the exterior piece is an open surface and should be continued inside the cylinder $x^2 + y^2 = a^2$ by means of an interior piece of $\mathcal C$. The two situations are shown in Fig.~\ref{fig2} for $D=5$. Looking at the figures, it is suspected that the interior piece of the trapped surface could hide a hole, such that the trapped surface would have torus topology.

\subsection{Interior trapped surface (for $D=5$)}

The interior piece of the trapped surface makes sense only for $D=5$, given that in other dimensions there is a singularity which takes up the whole disk $x^2 + y^2 \leq a^2$, $R =0$. 

Let $\Psi_-(\bar \rho,\bar R)$ be now the function for the piece of the trapped surface inside of  $x^2 + y^2 = a^2$. Parameterizing the surface $\mathcal C$ inside $x^2 + y^2 = a^2$ as $\bar \rho =\bar \rho(\bar R)$, the solution to the first equation in~\eqref{k54}, with the first boundary condition, is
\begin{equation}\label{k65}
\Psi_-(\bar \rho,\bar R) = \Phi_-(\bar \rho,\bar R) - \Phi_-(\bar \rho(\bar R), \rho) \,.
\end{equation}
Since $\Psi_-(\bar \rho,\bar R)|_{\mathcal C} =0$, we have that $\partial_{\bar R}\Psi_-(\bar \rho(\bar R),\bar R)=0$. Thus, the second boundary condition takes the appearance
\begin{equation}\label{k66}
 \partial_{\bar \rho}\Psi_-(\bar \rho(\bar R),\bar R) = -2 a^2,
\end{equation}
where we have used the fact that $\partial_\eta\Psi_-|_{\mathcal C} <0$ becayse $\Psi_-$ is defined such that $\Psi_- > 0$ inside $\mathcal C$ and $\Psi_-=0$ in $\mathcal C$. Using now the expression of Eq.~\eqref{k65}, one arrives at
\begin{align}\label{k67}
&\partial_{\bar \rho}\Psi_-(\bar \rho(\bar R),\bar R) = \partial_{\bar \rho}\Phi_-(\bar \rho^2(\bar R), R^2) = \nonumber \\
&
\frac{\mu \sqrt{\pi}}{a^{D-4}}\sum_{n=1}^{\infty}\frac{(-1)^n 2n \bar \rho(\bar R)^{2n-1}}{2^n n!\bar  R^{D-5} (1 + \bar R^2)^{\frac{4n+1}{2}}} \sum_{k=0}^{n}\binom{n}{k} P_n^{(k)}(D) \bar R^{2k} \nonumber \\
&
\times \frac{\Gamma\left(\frac{D-4+4n-2k}{2}\right)}{\Gamma\left(\frac{D-3+4n-2k}{2}\right)} {\ }_2F_1 \left(\frac{1}{2}, \frac{D-5-2k}{2}; \frac{D-3+4n-2k}{2}; -\frac{1}{\bar R^2}\right) \nonumber \\
&
+ \frac{2\mu \sqrt{\pi}}{a^{D-4}}\sum_{n=1}^{\infty} \frac{(-1)^n 2n \bar \rho(\bar R)^{2n-1} }{2^{n} n! \bar R^{D-3} (1 + \bar R^2)^{\frac{4n-1}{2}}} \sum_{k=0}^{n}\binom{n}{k} P_n^{(k)}(D+2) \nonumber \\ 
&
\times\bar  R^{2k}\left[ \frac{\Gamma\left(\frac{D-4+4n-2k}{2}\right)}{\Gamma\left(\frac{D-3+4n-2k}{2}\right)} {\ }_2F_1 \left(\frac{1}{2}, \frac{D-3-2k}{2}; \frac{D-3+4n-2k}{2}; -\frac{1}{\bar R^2}\right) \right. \nonumber \\
&
\left.- \frac{\Gamma\left(\frac{D-2+4n-2k}{2}\right)}{\Gamma\left(\frac{D-1+4n-2k}{2}\right)} {\ }_2F_1 \left(\frac{1}{2}, \frac{D-3-2k}{2}; \frac{D-1+4n-2k}{2}; -\frac{1}{\bar R^2}\right) \right] = -2 a^2 \,. 
\end{align}

Actually, Eq.~\eqref{k67} is a very hard algebraic equation to solve; but suppose again a power series expansion
\begin{equation}\label{k68}
\bar\rho(\bar R) = \bar \rho_0 + \bar \rho_2 \bar R^2 + \ldots,
\end{equation}
and solve order by order in $\bar R$. We are mainly interested on the leading order $\bar \rho_0$, since a solution $\bar \rho \neq 0$ would imply a hole in the trapped surface. For $D=5$, we have 
\begin{align}\label{k69}
& 
\frac{\mu}{a}\sum_{n=1}^{\infty}\frac{(-1)^n 2n \bar \rho_0^{2n-1}}{2^n n!} \sum_{k=0}^{n}\binom{n}{k} \frac{P_n^{(k)}(5)}{(2n)!} \nonumber \\
&
\times \Gamma\left(\frac{1+4n-2k}{2}\right)\Gamma\left(\frac{2k+1}{2}\right) \nonumber \\
&
+ \frac{2\mu}{a}\sum_{n=1}^{\infty} \frac{(-1)^n 2n \bar \rho_0^{2n-1} }{2^{n} n!} \sum_{k=0}^{n}\binom{n}{k} \frac{P_n^{(k)}(7)}{(2n)!} \nonumber \\ 
&
\times\Gamma\left(\frac{1+4n-2k}{2}\right)\Gamma\left(\frac{2k-1}{2}\right) \left[ 2n - \frac{1+4n-2k}{2+4n-2k}\frac{2k-1}{2}\right]= -2 a^2 \,. 
\end{align}
For $\bar \rho_0 \ll 1$, which is clearly fulfilled because we are dealing with the inside of the cylinder $x^2 + y^2 =a^2$, the equation above reduces to, 
\begin{equation}\label{k70}
\frac{\mu}{a}\frac{11\pi}{2}\rho_0  + \mathcal O(\rho_0^2)= 2 a^2 \,.
\end{equation}
Thus, at first order,
\begin{equation}\label{k71}
\rho_0 = \frac{4}{11\pi} \frac{a^3}{\mu} \,.
\end{equation}
This result shows that, for $D=5$, the trapped surface has topology $\mathbb S_1 \times\mathbb S_1 \times \mathbb R$ for $a\neq 0$, as it was previously suspected. 

Note that, for $\rho_0 = a$, we are in a extreme situation where the interior piece of $\mathcal C$ contains the ring singularity and is over the boundary $x^2 + y^2 = a^2$ of the interior region. From Eq.~\eqref{k71}, this happens when $a$ reaches the value
\begin{equation}\label{k72}
 a_1^2 = \frac{11\pi}{4} \mu \simeq 8.63 \mu\,.
\end{equation}
Then we can assure that there is no Penrose trapped surface for $|a|>a_1$. Since, from Eq.~\eqref{k64}, $a_0 < a_1$, we can take $a_0$ as an upper bound for the Kerr parameter that makes conditional the formation of the Penrose trapped surface in the collision.


\section{Conclusions}\label{sec7}

We have computed the gravitational shock wave geometry which arises from extremely boosting  the Kerr-like line element in various dimensions. The boost is done in the direction of the angular momentum and such that the ring singularity is preserved after the lightlike limit.  We have found a perturbative method which enable us to compute the profile function of the shock wave inside and outside the region bounded by the ring singularity. Then, we have argued that only for $D=5$ dimensions a complete solution for the profile function, covering inside and outside the ring singularity, is possible.

Although Kerr-like spacetimes have angular momentum, after performing the lightlike limit the property of a classical angular momentum is lost. However, the axis of symmetry of the Kerr-like spacetime is inherited in the shock wave geometry through the survival of the Kerr parameter $a$ in the lightlike limit. This fact makes strong conditions over the result of a head-on collision of two shock waves of the type considered here. 

The Penrose trapped surface formation has been considered over the head-on collision of two identical shock waves. Since for $D\ne 5$ the shock wave geometry diverges in extended regions, the Penrose trapped surface only appears for $D=5$ dimensions. For $D=5$ we have found that, if the Penrose surface forms, it has non-trivial topology $\mathbb R\times \mathbb S_1 \times \mathbb S_1$. 

Even for $D=5$ the Penrose trapped surface depends on the values of $\mu$ (relativistic energy) and $a$ (Kerr-like parameter), and could not be produced in the collision if the values of $\mu$ and $a$ are not appropriate. In this sense, we have found an upper bound $a_0^2\simeq  0.73 \mu$ for the formation of the Penrose surface. However, it should be noticed that the boundary $\mathcal C$ for $a$ = $a_0$ crosses the region where the convergence of Eq.~\eqref{k48} must be carefully analyzed. Therefore, a better upper bound could be found from an exhaustive study of the shock wave geometry near the ring singularity. 


\begin{acknowledgments}
This work has been partially funded by 
the Escuela Polit\'ecnica Nacional under projects PII-DFIS-2022-01 and PIM 19-01; 
the Ministerio Espa\~nol de Ciencia e Innovaci\'on under grant No. PID2019-107844GB-C22; the Junta de Andaluc\'ia under contract Nos. Operativo FEDER Andaluc\'ia 2014-2020 UHU-1264517, P18-FR-5057 and also PAIDI FQM-370.
\end{acknowledgments}


\bibliography{KerrShockWaves}

\end{document}